\documentclass[12pt]{article}

\usepackage{amssymb}
\usepackage{amsmath}
\usepackage{amsthm}

\newcommand{\dd}{\frac{d^2}{dx^2}}

\newcommand{\ep}[1]{\mathrm{e}^{#1}}
\newcommand{\I}{\mathrm{i}}
\newcommand{\tr}{\operatorname{tr}}
\newcommand{\Tr}{\operatorname{Tr}}
\newcommand{\sgn}{\operatorname{sgn}}
\newcommand{\DEF}{\mathrel{\mathop:}=}
\newcommand{\GL}{\mathrm{GL}}
\newcommand{\gl}{\mathrm{gl}}
\newtheorem{thm}{Theorem}
\newtheorem{lm}[thm]{Lemma}

\title{Equivalence of topological and scattering approaches to quantum
pumping}
\author{G. Br\"aunlich, G.M. Graf, G. Ortelli\\
\normalsize\it Theoretische Physik,
ETH-Z\"urich, CH--8093 Z\"urich}

\begin{document}

\maketitle
\begin{abstract}
The Schr\"odinger equation with a potential periodically varying in time 
is used to model adiabatic quantum pumps. The systems considered may be 
either infinitely extended and gapped or finite and connected to gapless
leads. Correspondingly, two descriptions of the transported charge, one 
relating to a Chern number and the other to a scattering matrix, have been 
available for some time. Here we generalize the first one and establish its 
equivalence to the second.

\end{abstract}

\maketitle

\section{Introduction}
Quantum pumps are driven devices connected to leads kept at a same voltage. 
Two descriptions of
charge transport are available for pumps depending on time periodically and 
adiabatically. One has been proposed by Thouless \cite{Th} (see also 
\cite{ThN}), the other by B\"uttiker et al. \cite{BTP} (see also \cite{B}). 
We shall refer to them as the topological, resp. the scattering 
approaches and denote by $\langle Q_T\rangle$, resp. 
$\langle Q_{BPT}\rangle$ the charges transported during a cycle.
Each one depends on a different idealization of the devices. In the first
proposal the model is a non-interacting Fermi gas, infinitely extended in one
dimension with the Fermi energy lying in a gap. The charge transported within
a period appears as a Chern number, indicating that it is quantized. In the 
second approach the device is viewed as a compact object connected to leads
containing free, gapless Fermi gases. Here, the transported charge is 
expressed in terms of the scattering matrix at Fermi energy and is quantized
in special cases only. 

At first sight charge transport is accounted for in rather different, if not
opposing, ways: The spatial extent of the two devices is infinite, resp. 
finite, reflecting a microscopic, resp. macroscopic, perspective; 
more strikingly, in the first case transport is attributed to 
energies way below the Fermi energy, which lies in a spectral gap, while in 
the second the scattering matrix matters only at Fermi energy. In physical
terms, the first description applies to insulators, the second
to conductors, at least seemingly so.

Yet, the two points of view are mathematically related. This has been shown 
in \cite{GO} for the
simpler case of a single channel, modeled as a real line, and of a potential
which is periodic also in space. A comparison becomes possible after truncating
the potential to finitely many periods, while the rest of the line gives raise
to the leads. Then the spectral gap closes and the model becomes amenable to 
the scattering approach. There, the conditions for quantized 
transport are attained in the limit of many periods, and quantitative agreement
between the two approaches was established. 

In this article we generalize the equivalence result in two ways, thereby
extending it to the natural setting of both approaches. First, the requirement
of spatial periodicity \cite{Th} is dropped. Such a situation was considered in
\cite{ThN}, though by approximating a general (e.g. quasi-periodic)
potential by a
sequence of periodic ones with increasing periods. Only the approximants were
associated to fiber bundles, based on the corresponding 
Brillouin zones. 
Here we propose a bundle and hence a Chern number applying directly 
to the infinite, non-periodic system. Second, we extend the correspondence 
\cite{GO} to a multi-channel setting.

As far as we know, the earliest statement concerning the equivalence is found 
in \cite{Chern}, though only for a particular, exactly solvable, periodic, 
tight binding Hamiltonian. On more general terms we note that, albeit the 
topological approach predates 
the scattering approach, several ideas underlying the equivalence can be 
traced back to \cite{Th}.
Experimental work which is thematically related is described e.g. in 
\cite{Shilton, Leek, Blu}.

In Section~\ref{section:2} we state the results for charge transport based
on the two approaches separately, and formulate the comparison, which is the 
main result, as Theorem~\ref{thm:2}.
In Section~\ref{section:3} we describe the relevant fiber bundle, while 
Section~\ref{section:4} is devoted to proofs. An appendix provides a result in
adiabatic perturbation theory.

\section{Main results}\label{section:2} 
We begin by describing the topological approach \cite{Th} in the case of 
$n$ channels. The Hamiltonian, acting on $L^2(\mathbb{R}_x, \mathbb{C}^n)$, is
\begin{equation} \label{eq:uno}
  H(s) = -\frac{d^2}{dx^2} + V(x, s) \,,
\end{equation}
where the potential $V=V(x, s)$ takes values in the
$n\times n$ matrices, $M_n(\mathbb{C})$, is Hermitian, $V=V^*$, and periodic
in time, $V(x,s + 2\pi)= V(x, s)$. For simplicity, let 
$V(\cdot,s)\in L^\infty(\mathbb{R}_x, M_n(\mathbb{C}))$ with $C^1$-dependence 
on $s\in S^1\DEF\mathbb{R}/2\pi\mathbb{Z}$. Then, for any $z\in\rho(H(s))$ 
in the resolvent set, the 
Schr\"odinger equation $H(s)\varphi=z\varphi$ is in the limit-point case at 
$x=+\infty$ (see \cite{L} or \cite{CG, LM}), meaning that as an ordinary 
differential equation it has $n$ linearly independent solutions which are 
square-integrable at
$x=+\infty$. We may thus introduce a family of sets,
parametrized by $z\in\rho(H(s))$ and 
$s\in S^1$, consisting of 
matrix-valued solutions $\psi(x)\in M_n (\mathbb{C})$ of the 
Schr\"odinger equation
\begin{equation} \label{eq:psi}
    -\psi'' (x) + V(x,s) \psi(x) = z \psi(x)  \,, 
\end{equation}
which are regular in the sense that for any $x\in\mathbb{R}$
\begin{equation}\label{eq:regular}
\psi(x)a=0,\, \psi'(x)a=0
\;
\Rightarrow\; a= 0\,, \qquad (a\in\mathbb{C}^n)\,.
\end{equation}
It is:
\begin{equation}\label{eq:psi+}
    S^+_{(z,s)} = \{ \psi_+ | \psi_+ \textrm{ is a regular solution of } 
(\ref{eq:psi}), L^2 \textrm{ at } x=+\infty \} \, .
\end{equation}
As a matter of fact such solutions tend to zero pointwise as $x\to +\infty$, 
together with their first derivatives. Similarly, solutions 
$\tilde\psi(x)\in M_n (\mathbb{C})$ of the adjoint equation
\begin{equation}\label{eq:chi}
    -\tilde\psi '' (x) + \tilde\psi(x) V(x,s) = z \tilde\psi(x) 
\end{equation}
act on row vectors $a\in\mathbb{C}^n$ as $a\tilde\psi(x)$, and we set
\begin{equation*}
\tilde{S}^-_{(z,s)} = \{ \tilde\psi_- | \tilde\psi_- \textrm{ is regular solution of } (\ref{eq:chi}), L^2 \textrm{ at } x=-\infty \} \, .
\end{equation*}
For later use we also introduce the families 
$S^-_{(z,s)}$, $\tilde{S}^+_{(z,s)}$ of solutions
to (\ref{eq:psi}), resp. (\ref{eq:chi}) decaying at the opposite ends.
For any two differentiable functions 
$\psi,\,\tilde\psi:\mathbb{R}\to M_n(\mathbb{C})$ we define the Wronskian
\begin{equation}
\label{eq:wr} 
W(\tilde\psi, \psi ; x) = 
\tilde\psi (x) \psi '(x) - \tilde\psi ' (x) \psi (x) \in M_n(\mathbb{C})\, .
\end{equation}
It is independent of $x$ if $\psi$ and $\tilde\psi$ are solutions 
of (\ref{eq:psi}), resp. of (\ref{eq:chi}), in which case it is simply 
denoted as
$W(\tilde\psi_-, \psi_+)$. As will also be shown later,
$\det W(\tilde\psi_-, \psi_+)\neq 0$ for $\psi_+\in S^+_{(z,s)}$, 
$\tilde\psi_-\in \tilde{S}^-_{(z,s)} $.
We observe that $S^+_{(z,s)}$ carries a transitive right action of 
$\GL(n)\ni T$,
\begin{equation}
\label{eq:ra}
    \psi_+ (x) \mapsto \psi_+ (x) T \, , 
\end{equation}
while $\tilde{S}^-_{(z,s)}$ carries a left action,
\begin{equation*}
    \tilde\psi_- (x) \mapsto T \tilde\psi_- (x) \, .
\end{equation*}
We thus have a bijective relation between $\psi_+\in S^+_{(z,s)}$ and 
$\tilde\psi_-\in \tilde{S}^-_{(z,s)}$ such that
\begin{equation}
    W(\tilde\psi_-, \psi_+ ) =1 \, .
\label{eq:nrmlzt}
\end{equation}
We assume that the Fermi energy $\mu>0$ lies in a spectral gap at
all times $s$: 
\begin{equation}
    \mu\in \rho(H(s))\,.
\label{eq:gap}
\end{equation}
Let $P_0(s)$ be the spectral projection of $H(s)$ up to the Fermi energy and 
$U_\varepsilon (s, s_0)$ be the propagator for the non-autonomous 
Hamiltonian $H(\varepsilon t)$, where $s=\varepsilon t$. In the Appendix 
we prove, in the smooth case, 
\begin{equation}\label{eq:Pzero}
    U_\varepsilon (s, s_0) (P_0(s_0)+ \varepsilon P_1(s_0)) U_\varepsilon (s, s_0)^* = 
P_0 (s) + \varepsilon P_1 (s) + O (\varepsilon^2 )\,, \qquad 
(\varepsilon \rightarrow 0)
\end{equation}
with
\begin{equation}\label{eq:Puno}
    P_1 (s) = -\frac{1}{2\pi} \oint_\gamma R(z,s) \dot{R} (z,s) dz\,,
\end{equation}
where $R(z,s) = ( H(s) - z)^{-1}$ and $\gamma$ is a complex contour 
encircling the part of the spectrum of 
$H(s)$ lying below $\mu$ and $\dot{} = \partial / \partial s$. 
Eq.~(\ref{eq:Pzero}) is the 1-particle density matrix which has 
evolved from that of the Fermi sea, $P_0(s_0)$,  after a gentle start of the
pump. In fact such a start may 
be obtained from (\ref{eq:uno}) by means of a smooth substitution 
$s'\mapsto s$ with 
$s'\mapsto s_0$, ($s\le s_0$), and $s'=s$, ($s'$ large). Then, in the new
variable, $P_1(s_0)=0$ by (\ref{eq:Puno}). 

The current across a fiducial point $x=x_0$ is the rate of change of the 
charge contained in $x>x_0$ and hence given by the
operator $I=\I[H(s), \theta(x-x_0)]$, which is independent of $s$. The charge 
transported in a cycle (of duration $2\pi\varepsilon^{-1}$) is, in 
expectation value and in the adiabatic limit, given as
\begin{equation}\label{eq:QT}
\langle Q_T\rangle\DEF\oint \Tr(IP_1(s))ds\,,
\end{equation}
because of $dt=\varepsilon^{-1}ds$, with $\Tr$ denoting the trace on
$L^2(\mathbb{R}_x, \mathbb{C}^n)$. This definition rests on the fact that the
leading contribution from persistent currents, 
$\varepsilon^{-1}\oint\Tr(IP_0(s))ds$, which is potentially divergent in 
the limit, actually vanishes. If $V$ were real, this would follow 
trivially from time reversal invariance; however our hypothesis does not imply
this, except for $n=1$, and we shall argue otherwise.

The result of \cite{Th}, generalized as described in the Introduction, 
is part (ii) of the following theorem.
\begin{thm} 
\label{thm:1}
Assume (\ref{eq:gap}). Then
\begin{enumerate}
\item
\begin{equation*}
\Tr(IP_0(s))=0\,.
\end{equation*}
\item
\begin{equation}
\langle Q_T\rangle=\frac{\I}{2\pi}\oint_\gamma dz \oint_{S^1} ds\,
\tr \bigl( W(\frac{\partial \tilde\psi_-}{\partial z} , \frac{\partial \psi_+}{\partial s};x_0) - W( \frac{\partial \tilde\psi_-}{\partial s} , \frac{\partial \psi_+}{\partial z};x_0) \bigr) \,,
\label{eq:th}
\end{equation}
where $\tr$ denotes the matrix trace and the solutions 
$\psi_+\in S^+_{(z,s)}$, 
$\tilde\psi_-\in \tilde{S}^-_{(z,s)}$ satisfying (\ref{eq:nrmlzt}) are 
locally smooth in $(z,s)$. Except for these conditions, the trace is
independent of $\psi_+$, $\tilde\psi_-$, and the integral is it of $x_0$, 
too. Moreover, the
r.h.s. is the first Chern number of a bundle described in 
Section~\ref{section:3}.
\end{enumerate}
\end{thm} 

We next present the scattering description \cite{BTP} of charge
transport. Consider again the Hamiltonian (\ref{eq:uno}), but now with $V$ of
compact support in $x$. As a result, (\ref{eq:gap}) fails:
\begin{equation}
    \mu\in \sigma(H(s))
\label{eq:nogap}
\end{equation}
for all $s$. We may thus introduce the scattering matrix $S(s)$ at Fermi energy
$\mu>0$, 
\begin{equation*}
  S(s)=\begin{pmatrix}R&T'\\T&R'\end{pmatrix}\,,
\end{equation*}
where the blocks are $n\times n$ matrices determined by the asymptotic 
behavior of solutions of (\ref{eq:psi}) with $z=\mu$. More precisely,
$R$ and $T$ are defined in terms of a plane wave incident from the left,
\begin{equation}\label{eq:incident}
 \psi(x)=\begin{cases}
1\ep{\I kx}+R\ep{-\I kx}\,,&(x<-r)\,,\\
T\ep{\I kx}\,,&(x>r)\,,
\end{cases}
\end{equation}
with $r>0$ large enough and $k=\sqrt{\mu}$. Similarly $R'$ and $T'$ 
are defined in terms of a wave incident from the right.

The charge emitted from all channels of the left lead together, in a cycle 
and in the adiabatic limit, is \cite{BTP}
\begin{equation}
\langle Q_{BPT}\rangle=\frac{1}{2\pi\I}\oint\tr((dS)S^*P)\,,
\label{eq:bpt}
\end{equation}
where $dS=(dS/ds)ds$ and 
$P=\bigl(\begin{smallmatrix} 1&0\\0&0\end{smallmatrix}\bigr)$
is the projection onto the left channels. For the same situation the variance 
is \cite{ILL, AEGS}
\begin{equation*}
\langle\langle Q_{BPT}^2\rangle\rangle=\frac{1}{(2\pi)^2}\int_{-\infty}^\infty
ds\oint ds'\frac{\tr[(S^*(s)PS(s)-S^*(s')PS(s'))^2]}{\sin^2{(s-s')}}\,.
\end{equation*}
In general, and in contrast to (\ref{eq:th}), $\langle Q_{BPT}\rangle$ is not
an integer. However, $\langle\langle Q_{BPT}^2\rangle\rangle$ vanishes 
iff the time dependence of $S$ is of the form 
\begin{equation}
S(s)=\begin{pmatrix}U_1(s)&0\\0&U_2(s)\end{pmatrix}S_0
\label{eq:qntz}
\end{equation}
with $U_j(s)$ ($j=1,2$) and $S_0$ unitary matrices of order $n$, resp. $2n$.
In this case $\langle Q_{BPT}\rangle$ is an integer,
\begin{equation*}
\langle Q_{BPT}\rangle=\frac{1}{2\pi\I}\oint\tr((dU_1)U_1^*)
=\frac{1}{2\pi\I}\oint d\log\det U_1\,,
\end{equation*}
given as the winding number of $\det U_1$.\\

We do not give here the definition of $\langle Q_{BPT}\rangle$ which makes
(\ref{eq:bpt}) a theorem \cite{AEGSS}. Rather we focus on the relation between 
Eqs.~(\ref{eq:th}) and (\ref{eq:bpt}). To this end we truncate the potential 
to a
finite interval, $V(x,s)\chi_{[0,L]}(x)$, and denote its scattering matrix by 
$S_L(s)$. In the limit $L\to\infty$ the original physical situation is
recovered and the two approaches agree, as stated in the following result. 
\begin{thm}\mbox{}
\label{thm:2} Assume (\ref{eq:gap}) for the infinite system.
\begin{enumerate}
\item
The scattering matrix $S_L(s)$ at Fermi energy $\mu$ has a limit of the form 
\begin{equation}
\label{eq:cinque}
\lim_{L\to\infty}  S_L(s)=\begin{pmatrix}R(s)&0\\0&R'(s)\end{pmatrix}\,.
\end{equation}
In particular, the condition 
(\ref{eq:qntz}) for quantization of $\langle Q_{BPT}\rangle$ is attained 
in the limit.
\item
The winding number of $\det R(s)$ equals the Chern number on the r.h.s of 
Eq.~(\ref{eq:th}). In physical terms, 
\begin{equation}
\label{eq:equiv}
\langle Q_{BPT}\rangle=\langle Q_T\rangle\,.
\end{equation}
\end{enumerate}
\end{thm}
We conclude this section by summarizing the idea of the 
proof of (\ref{eq:equiv}). We may assume that the contour $\gamma$ in 
Eqs.~(\ref{eq:Puno}, \ref{eq:th}) crosses the real axis just twice, once 
below the spectrum and once at Fermi energy $\mu$. The torus of integration in 
(\ref{eq:th}), which is denoted by $\mathbb{T}=\gamma\times S^1$, is the base
space of a bundle which will admit a global section except at isolated points
along the line $\{\mu\}\times S^1\subset\mathbb{T}$. Using Stokes' theorem
its Chern number can be expressed in terms of solutions of the Schr\"odinger
equation at Fermi energy and, in turn, of the scattering matrix 
(\ref{eq:cinque}). The main steps are given in more detail in the following
lemma. There the r.h.s. of eq.~(\ref{eq:th}) is denoted by $C$, and $x_0$ 
is fixed. The orientation of the torus is the natural one, $d\gamma\wedge ds$. 
\begin{lm}
\label{mainlm}
\begin{enumerate}
\item
Any point $(z_*, s_*)\in \mathbb{T}$ where $\det\psi_+ (x_0)=0$ for some (and 
hence all) $\psi_+\in S^+_{(z_*,s_*)}$ has $z_*=\mu$. For a dense set of
potentials $V=V^*$, the points 
$s_*$ are isolated in $S^1$ and $0$ is a simple 
eigenvalue of $\psi_+ (x_0)$; moreover, 
\begin{equation}\label{eq:reg}
\det\psi'_+ (x_0)\neq0\,.
\end{equation}
Density is meant w.r.t. the topology of the class of potentials 
specified below (\ref{eq:uno}).
\item
Let $\psi_{(z,s)}\in S^+_{(z,s)}$ be a section defined in a neighborhood in 
$\mathbb{C}\times S^1\supset \mathbb{T}$ of any of the above points 
$(z_*=\mu, s_*)$, which is analytic in $z$. Then the family of matrices
$L(z,s)=\psi'_{(\bar z,s)}(x_0)^*\psi_{(z,s)}(x_0)$ has the 
reflection property
\begin{equation}\label{eq:refl}
L(z,s)=L(\bar z,s)^*\,.
\end{equation}
Its eigenvalues are real for real $z$. There is a single eigenvalue branch 
$\lambda(z,s)$ vanishing to first order at $(\mu, s_*)$. Its winding 
number there is 
\begin{equation*}
w_{s_*}=-\sgn\bigl(\frac{\partial\lambda}{\partial z}\frac{\partial \lambda}
{\partial s}\bigr)\Big|_{(z=\mu,s= s_*)}\,.
\end{equation*}
\item
\begin{equation*}
C=-\sum_{s_*}w_{s_*}\,.
\end{equation*}
\item At any of the points $(\mu,s_*)$ we have
\begin{equation*}
\frac{\partial\lambda}{\partial z}<0\,.
\end{equation*}
\item
The unitary matrix $R(s)$ has eigenvalue $-1$ iff $\det\psi_{\mu,s}(0)=0$. 
More precisely, as $s$ increases past $s_*$, an eigenvalue of $R$ crosses $-1$ 
counterclockwise if 
\begin{equation*}
\frac{\partial \lambda}{\partial s}\Big|_{(z=\mu,s= s^*)}<0\,. 
\end{equation*}
\end{enumerate}
\end{lm}
As a result, 
$C=-\sum_{s_*}\sgn\bigl(\partial \lambda/\partial s\bigr)|_{(z=\mu,s= s_*)}$ 
is the number of eigenvalue crossings of $R(s)$ past $-1$, 
i.e., the winding number of $\det R$. Actually the equality is first 
established if the conditions on the potential of part (i) are satisfied, but 
the conclusion, Eq.~(\ref{eq:equiv}), extends by density.

\section{A fiber bundle}\label{section:3} 

We describe the bundle $P$ and the connection underlying Eq.~(\ref{eq:th}).
Let $\mathcal{C} = C^1 \bigl(\mathbb{R}, M_n (\mathbb{C}) \bigr)$ be the 
space of matrix valued $C^1$-functions on $\mathbb{R}$. Let $\pi:
P\to \mathbb{T}$ be the subbundle of $\mathbb{T}\times\mathcal{C}$
with base $\mathbb{T}=\gamma\times S^1$
and fibers $S^+_{(z,s)}\subset\mathcal{C}$:
\begin{equation*}
P=\{((z,s),\psi)\in \mathbb{T}\times \mathcal{C}\mid \psi\in S^+_{(z,s)}\}\,.
\end{equation*}
It is a principal bundle w.r.t. 
the right action (\ref{eq:ra}) of $\GL(n)$. This includes that
$\GL(n)$ is its structure group. Indeed, for any sufficiently small
open set $U\subset\mathbb{T}$ there is $x\in\mathbb{R}$ with 
\begin{equation*}
\det\psi_+(x) \neq 0
\end{equation*}
for all $\psi_+\in S^+_{(z,s)}$ and $(z,s)\in U$, see Lemma~\ref{lm:prlm}
below. This provides a local trivialization $\phi$ with
\begin{equation*} 
\phi^{-1}:\pi^{-1}(U)\to U\times \GL(n)\,,\quad
\psi_+\mapsto(z,s,\psi_+(x))\,.
\end{equation*}
The transition function $\phi_2^{-1}\circ\phi_1: \GL(n)\to\GL(n)$ is
multiplication from the left by the matrix
$\psi_+(x_2)\psi_+(x_1)^{-1}$, which is clearly independent of
$\psi_+\in S^+_{(z,s)}$ and belongs to $\GL(n)$.

We will give an explicit expression for the
Chern number $C$ of $P$, which differs somewhat from that used in
\cite{Th}. We recall that
\begin{equation}\label{eq:C}
    C = \frac{\I}{2\pi} \int_\mathbb{T} \tr \mathcal{F} \,,
\end{equation}
where $\mathcal{F} = D \mathcal{A}$ is the curvature of any connection
$\mathcal{A}$ on $P$. We recall that $\tr \mathcal{F}$ defines a 2-form on 
$\mathbb{T}$, and not just on $P$; for any two connections, $\mathcal{A}$ 
and $\mathcal{A}'$, the same is true for the 1-form 
$\tr(\mathcal{A} - \mathcal{A}')$, whence $C$ is independent of the choice of 
connection. We consider connections of the following form. Let  
$B:  \mathcal{C} \times \mathcal{C} \rightarrow M_n (\mathbb{C})$ be
a bilinear form on $\mathcal{C}$ satisfying
\begin{align}
    B(\tilde\psi , \psi T) &= B(\tilde\psi , \psi)T\,, \label{eq:B1} \\
    B(T \tilde\psi , \psi) &= T B(\tilde\psi ,\psi) \label{eq:B2}
\end{align}
($\tilde\psi$, $\psi \in \mathcal{C}$, $T \in \GL(n)$). Moreover we
assume that its restriction 
\begin{equation}\label{eq:B3}
    B: \tilde{S}^-_{(z,s)} \times S^+_{(z,s)} \rightarrow \GL(n)
\end{equation}
takes values $B(\tilde\psi_- , \psi_+)$ in the regular matrices 
(as shown below, an example is (\ref{eq:wr})). We may then consider the 
$\gl(n)$-valued 1-form on $P$
\begin{equation*}
    \mathcal{A}_{\psi_+} (\delta \psi_+) = B(\tilde\psi_- ,
\psi_+)^{-1} B(\tilde\psi_- , \delta \psi_+ ) \, ,\qquad(\delta
\psi_+\in TP)\, ,
\end{equation*}
which is well-defined being independent of the choice of $\tilde\psi_-
\in \tilde{S}^-_{(z,s)}$ by (\ref{eq:B2}). It is a connection on $P$
since it enjoys the defining properties 
\begin{align*}
    \mathcal{A}_{\psi_+} (\psi_+ t) &= t \, , \qquad (t \in \gl(n))\, , \\
    \mathcal{A}_{\psi_+ T} (\delta \psi_+ T) &= T^{-1} \mathcal{A}_{\psi_+} (\delta \psi_+) T \, , \qquad (T \in \GL(n))
\end{align*}
by (\ref{eq:B1}). Given $\psi_+ \in S^+_{(z,s)}$ there is a unique
$\tilde\psi_- \in \tilde{S}^-_{(z,s)}$ such that $B(\tilde\psi_- ,
\psi_+) = 1$, as can again be seen from (\ref{eq:B2}). Then
$\mathcal{A} = B(\tilde\psi_- , \delta \psi_+)$ and the trace of its
curvature is 
\begin{equation*}
  \tr \mathcal{F} = \tr \bigl( B(\frac{\partial\tilde\psi_-}{\partial z} , \frac{\partial \psi_+}{\partial s}) - B( \frac{\partial \tilde\psi_-}{\partial s} , \frac{\partial \psi_+}{\partial z}) \bigr) dz\wedge ds\, .
\end{equation*}
We will use the bilinear
\begin{equation*}
    B(\tilde\psi , \psi) = W(\tilde\psi, \psi ; x) = \tilde\psi (x) \psi '(x) - \tilde\psi ' (x) \psi (x) \, ,
\end{equation*}
whose restriction (\ref{eq:B3}) is seen to be
independent of $x$ (though $\mathcal{A}$ may not be); then (\ref{eq:C}) 
coincides with the r.h.s. of
(\ref{eq:th}), as announced in Theorem~\ref{thm:1}. It remains to verify
$B(\tilde\psi_- , \psi_+) \in \GL(n)$. Any column vector solution
$\varphi (x)$ of (\ref{eq:psi}) is determined by $\varphi(0)$, 
$\varphi'(0) \in \mathbb{C}^n$. Similarly for any row vector $\tilde\varphi
(x)$ solving (\ref{eq:chi}). Their Wronskian  
\begin{equation}\label{eq:W}
    W(\tilde\varphi , \varphi) = \tilde\varphi (0) \varphi'(0) - 
\tilde\varphi'(0) \varphi (0) \, ,
\end{equation}
which now takes values in $\mathbb{C}$, clearly defines a
non-degenerate bilinear form on $\mathbb{C}^{2n}$. Given 
$\psi_\pm \in S^{\pm}_{(z,s)}$, any solution $\varphi$ can be expressed as
\begin{equation}
\label{eq:gs}
    \varphi (x) = \psi_+ (x) a_+ + \psi_- (x) a_-
\end{equation}
with $a_\pm \in \mathbb{C}^n$, and $\varphi \equiv 0$ iff $a_\pm = 0$;
similarly for $\tilde\varphi (x) = b_+ \tilde\psi_+ (x) + b_-
\tilde\psi_- (x)$. In terms of the coefficients $(b_+ , b_-)$, $(a_+ ,
a_-)$, the bilinear form (\ref{eq:W}) is given by the matrix  
\begin{equation*}
    \left( \begin{array}{cc}
                0 & W(\tilde\psi_+ , \psi_-) \\ W(\tilde\psi_- , \psi_+) & 0
            \end{array} \right) \, ,
\end{equation*}
since 
\begin{equation}\label{eq:nrmlzt3}
    W(\tilde\psi_\pm , \psi_\pm ) = \lim_{x \rightarrow \pm \infty} W
(\tilde\psi_\pm , \psi_\pm ; x) = 0\,.
\end{equation}
Hence $W(\tilde\psi_- , \psi_+)$ is regular. \\

\noindent {\bf Remark.} In \cite{Th} (and later in \cite{GO}) the case of a potential $V(x)$ of period $L$ was considered. In the case $n=1$ the bilinear used there was
\begin{equation*}
    B(\tilde\psi , \psi) = \int_0^L dx \, \tilde\psi (x) \psi (x) \, .
\end{equation*}
Non-degeneracy of (\ref{eq:B3}) amounts to
$\int_0^L dx \, \psi_- (x) \psi_+ (x) \neq 0 $,
where $\psi_-\in \tilde{S}^-_{(z,s)} = S^-_{(z,s)}$, $\psi_+\in S^+_{(z,s)}$ 
are unique up to non-zero multiples.

\section{Proofs}\label{section:4} 

Here we prove Theorems~\ref{thm:1} and~\ref{thm:2} stated in 
Section~\ref{section:2}. First however we 
should dwell on a little point of precision: The current, informally given as 
\begin{equation}
I=\I[H, \theta(x)]=-\I\bigl\{\frac{d}{dx},\delta(x)\bigr\}\,,
\label{eq:cu1}
\end{equation}
is not a well-defined operator on Hilbert space. (We suppressed $s$ from 
the notation and set $x_0=0$.) Instead, it should be understood as the map 
$D(H)\to D(H)^*$,
\begin{equation*}
I=\I(\gamma_1^*\gamma_0-\gamma_0^*\gamma_1)\,,
\end{equation*}
where $\gamma_0, \gamma_1:D(H)\to\mathbb{C}^n$ with $\gamma_0\psi=\psi(0)$,
$\gamma_1\psi=\psi'(0)$. Then (\ref{eq:cu1}) is replaced by
\begin{equation}
\I[R(z), \theta(x)]=-R(z)IR(z)\,,
\label{eq:cu2}
\end{equation}
which can be verified first as a quadratic form. This operator is of
trace class because $(p^2+1)^{-1}\gamma_i^*\gamma_i (p^2+1)^{-1}$ is. 

Given an operator $K:D(H)^*\to D(H)$ one may, pretending cyclicity, take
\begin{equation*}
\Tr(IK):=\I\tr(\gamma_0K\gamma_1^*-\gamma_1K\gamma_0^*)
\end{equation*}
as a definition. In fact, this is the trace of the finite rank operator $IK$
on the Banach space $D(H)^*$, see e.g.~\cite{Si}, Eq.~(10.2). It yields 
\begin{equation}
\Tr(IK):=\tr(-\I\partial_1K(0,0) +\I\partial_2K(0,0))\,,
\label{eq:cu3}
\end{equation}
where $K(x,y)$ is the integral kernel of $K$ and $\partial_1$ and $\partial_2$ 
indicate a derivative w.r.t. the first, resp. second argument. 
As a further motivation we note that expectation values of the
current are naturally written as $\Tr(P_0IP_0)$ and $\Tr(P_0IP_1+P_1IP_0)$ in
zeroth and first order in $\varepsilon$. Then 
\begin{equation}
\Tr(P_0IP_0)=\I\Tr\bigl(P_0(\gamma_1^*\gamma_0-\gamma_0^*\gamma_1)P_0\bigr)
=\I\tr(\gamma_0P_0\gamma_1^*-\gamma_1P_0\gamma_0^*)\,,
\label{eq:cu4}
\end{equation}
where cyclicity is now justified since $\gamma_i P_0$ is Hilbert-Schmidt; 
also, $P_0^2=P_0$ was used. Similarly,
\begin{equation*}
\Tr(P_0IP_1+P_1IP_0)=\I\tr(\gamma_0P_1\gamma_1^*-\gamma_1P_1\gamma_0^*)\,,
\end{equation*}
by $P_0P_1+P_1P_0=P_1$.\\

\noindent 
{\bf Proof of Theorem~\ref{thm:1}.}
i) The projection $P_0$ has the integral 
representation $P_0= -(2\pi \I)^{-1} \oint_\gamma R(z) \, dz$. Since 
$\oint_\gamma R(z)^2 \, dz=0$ we may replace $R(z)$ therein by 
$R(z)-R(z)^2H=-zR(z)^2$:
\begin{equation*}
P_0= \frac{1}{2\pi \I}  \oint_\gamma z R(z)^2 \, dz\,.
\end{equation*}
We then have, by (\ref{eq:cu4}, \ref{eq:cu2}),
\begin{align}
\Tr(P_0IP_0)&=\frac{1}{2\pi} \oint_\gamma z
\tr(\gamma_0R(z)^2\gamma_1^*-\gamma_1R(z)^2\gamma_0^*)\, dz\nonumber\\
&=\frac{1}{2\pi} \oint_\gamma z
\Tr\bigl(R(z)(\gamma_1^*\gamma_0-\gamma_0^*\gamma_1)R(z)\bigr) \, dz
=-\frac{1}{2\pi} \oint_\gamma z\Tr([R(z), \theta(x)]) \, dz\,,
\label{eq:cu5}
\end{align}
and, by $zR(z)=HR(z)-1$, also
$\Tr(P_0IP_0)=\I \Tr[HP_0,\theta]$. As the stationarity of $P_0$ suggests, 
the current is independent of $x_0$. In fact, upon replacing $\theta(x)$ by 
$\tilde\theta(x)=\theta(x-x_0)-\theta(x)$ both terms in 
$\Tr((HP_0)\tilde\theta-\tilde\theta(HP_0))$ are separately trace class,
whence the trace vanishes (\cite{Si}, Corollary 3.8). We next turn to 
(\ref{eq:cu5}):
The commutator $A=[R(z), \theta(x)]$ has integral kernel 
$A(x,y)=G(x,y)(\theta(y)-\theta(x))$, where $G(x, x') = R(z)(x,x')$ is 
the Green function. By the stated independence 
we may average over $x_0$ instead of setting it to $0$,
thus effectively smoothing $\theta$. We will see in 
(\ref{eq:grnfct}, \ref{eq:ids}) below that $G(x,y)$ is continuous. Thus
$A(x,x)=0$, implying
$\Tr(P_0IP_0)=0$. Alternatively the conclusion may be reached without
smoothing by resorting to Brislawn's theorem (\cite{Si}, Theorem A.2), 
according to
which $\Tr A=\int dx\, \tilde A(x,x)$, where $\tilde A(x,y)$ is the Lebesgue
value of $A(x,y)$. Here, $\tilde A(x,x)=0$.\\

\noindent
ii) By applying (\ref{eq:cu3}) to $K=R(z,s) \dot{R} (z,s)$ 
in (\ref{eq:QT}, \ref{eq:Puno}) we obtain for the transported charge
\begin{equation}
\label{eq:QT2}
\langle Q_T \rangle = \frac{\I}{2 \pi} \oint ds \oint_\gamma \! dz \int  dx \, \tr \bigl( \partial_1 G(0,x) \dot{G}(x,0) - G(0,x) \partial_2 \dot{G}(x,0) \bigr)\,.
\end{equation}
We claim that the Green function can be expressed as
\begin{equation}\label{eq:grnfct}
	G(x, x')= - \theta(x - x') \psi_+ (x) \tilde\psi_- (x') - \theta (x' - x) \psi_- (x) \tilde\psi_+ (x') \, ,
\end{equation}
where we complemented the locally smooth choice of
$\psi_+ \in S^+_{(z,s)}$, $\tilde{\psi}_- \in \tilde{S}^-_{(z,s)}$
satisfying (\ref{eq:nrmlzt}) by that of a pair 
$\tilde{\psi}_+ \in \tilde{S}^+_{(z,s)}$, $\psi_- \in S^-_{(z,s)}$ with 
\begin{equation}\label{eq:nrmlzt2}
	W( \tilde\psi_+ , \psi_- ) = -1 \, .
\end{equation}
Indeed, because of (\ref{eq:nrmlzt}, \ref{eq:nrmlzt2}) and of
(\ref{eq:nrmlzt3}) the general column solution (\ref{eq:gs}) has coefficients 
\begin{equation*}
a_\pm=\pm W( \tilde\psi_\mp,\varphi)=\pm\tilde\psi_\mp(y)\varphi'(y)
\mp\tilde\psi'_\pm(y)\varphi(y)\,.\\
\end{equation*}
By inserting this in (\ref{eq:gs}) and in its derivative w.r.t. $x$, and by
setting $y=x$, we conclude from the arbitrariness of $\varphi(x)$ and
$\varphi'(x)$ that 
\begin{align}
    \psi_+ (x) \tilde\psi_- (x) - \psi_- (x) \tilde\psi _+ (x) &= 0\,,
\label{eq:ids} \\
    \psi_+ (x) \tilde\psi_- ' (x) - \psi_- (x) \tilde\psi _+ ' (x) &=
    -1\,,
\nonumber\\
    \psi_+ ' (x) \tilde\psi_- (x) - \psi_- ' (x) \tilde\psi_+ (x) &= 1
    \, .
\nonumber
\end{align}
By means of these relations one verifies that $G$, as given by the r.h.s. 
of (\ref{eq:grnfct}), satisfies 
\begin{equation*}
\Bigl(-\frac{d^2}{dx^2}+V(x)-z\Bigr)G(x,x')=\delta(x-x')1\,;
\end{equation*}
together with $G(x,x')\to 0$, ($|x|\to\infty$), which exhibits it as the
Green function. We then apply (\ref{eq:grnfct}) in Eq.~(\ref{eq:QT2}): 
For $x\ge 0$ the integrand is
\begin{multline*}
\tr \bigl( \partial_1 G(0,x) \dot{G}(x,0) - G(0,x)
\partial_2 \dot{G}(x,0) \bigr)= \\
\tr \bigl( \psi_- ' (0) \tilde\psi_+ (x) 
( \dot{\psi}_+ (x) \tilde\psi_- (0) + \psi_+ (x)
\dot{\tilde\psi}_- (0) )
-  \psi_-(0) \tilde\psi_+ (x) 
( \dot{\psi}_+ (x) \tilde\psi_-' (0) + \psi_+ (x)
\dot{\tilde\psi}_-' (0) )\bigr)\\
= \tr \bigl( W(\dot{\tilde\psi}_- , \psi_- ) \, 
\tilde\psi_+ (x) \psi_+ (x)\bigr) \,,
\end{multline*}
where we used cyclicity of the trace and (\ref{eq:nrmlzt3}). Here and 
henceforth the Wronskian is evaluated at $x=0$, unless otherwise stated. 
Together
with a similar computation for $x\le 0$ we obtain
\begin{equation}\label{eq:QT3}
 \langle Q_T \rangle=  \frac{\I}{2 \pi} \oint \! ds \oint_\gamma \! dz \, \tr \bigl( W(\dot{\tilde\psi}_- , \psi_- ) \int_0^\infty \! dx \, \tilde\psi_+ (x) \psi_+ (x) +W( \dot{\tilde\psi}_+ , \psi_+ ) \int_{-\infty}^0 \! dx \, \tilde\psi_- (x) \psi_- (x) \bigr) \, .
\end{equation}
We maintain that the same expression is obtained from a computation of
$C$, the r.h.s. of (\ref{eq:th}). That calls for one of $\partial \psi_+ /
\partial z$, $\partial \tilde{\psi}_- / \partial z$. Differentiating 
(\ref{eq:psi}) w.r.t. $z$ we obtain
\begin{equation*}
    \Bigl( -\dd + V(x,s) -z \Bigr) 
\frac{\partial \psi_+}{\partial z} = \psi_+ \, ,
\end{equation*}
whose general solution with 
$\partial \psi_+ / \partial z \rightarrow 0$, ($x \rightarrow \infty$) is
\begin{equation}
    \frac{\partial \psi_+}{\partial z} (x) 
=\psi_+ (x)F_+(x)  - 
\psi_- (x) \int_x^{\infty} \tilde\psi_+ (x') \psi_+ (x') dx' \, ,\label{eq:dz1}
\end{equation}
where $F_+' (x) =dF_+/dx=-\tilde \psi_- (x)\psi_+ (x)$. Hence $F_+$ is 
determined up to an additive constant, which 
reflects the gauge freedom (\ref{eq:ra}) of
$\psi_+$. Eq.~(\ref{eq:dz1}) is verified by twice differentiating it w.r.t. 
$x$, the first derivative being
\begin{equation*}
    \frac{\partial \psi_+'}{\partial z} (x) 
=\psi_+ '(x)F_+(x) - 
\psi_-' (x) \int_x^{\infty} \tilde\psi_+ (x') \psi_+ (x') dx' \, ,
\end{equation*}
by using (\ref{eq:ids}). In the same way we find 
\begin{equation*}
    \frac{\partial \tilde\psi_-}{\partial z} (x) = F_-(x) \tilde\psi_- (x) - \Bigl( \int_{-\infty}^x \tilde\psi_- (x') \psi_- (x') dx' \Bigr) \tilde\psi_+ (x) \, ,
\end{equation*}
with $F_-' = -F_+'$. The arbitrariness of $F_\pm$ is constrained by 
(\ref{eq:nrmlzt}), which implies 
\begin{equation}\label{eq:dc1}
F_++F_-=0\,. 
\end{equation}
This is seen by differentiating the constraint w.r.t. $z$ and by using
\begin{align*}
W(\tilde\psi_-, \frac{\partial \psi_+}{\partial z} ; x)
&=W(\tilde\psi_-,\psi_+;x)F_+(x)
-W(\tilde\psi_-,\psi_-;x)\int_x^\infty \tilde\psi_+ (x) \psi_+ (x) \, dx 
=F_+(x)\,,\\
W(\frac{\partial \tilde\psi_-}{\partial z} , \psi_+ ; x)&=F_-(x)\,.
\end{align*}
Similarly, differentiating the constraint w.r.t. $s$ yields
\begin{equation}\label{eq:dc2}
    W( \dot{\tilde\psi}_- , \psi_+;x) + 
W( \tilde\psi_- , \dot{\psi}_+;x) = 0\,.
\end{equation}
We are now in position to compute $C$ and in particular
\begin{align*}
W(\frac{\partial \tilde\psi_-}{\partial s} , 
\frac{\partial \psi_+}{\partial z})
=& 
\dot{\tilde\psi}_- (0) \bigl( \psi_+ ' (0) F_+(0) -
\psi_- ' (0) \int_0^\infty  \tilde\psi_+(x) \psi_+ (x) \, dx \bigr) 
 \\ 
&  - \dot{\tilde\psi}_- ' (0) \bigl( \psi_+ (0) F_+(0) - \psi_- (0) \int_0^\infty \tilde\psi_+ (x) \psi_+ (x) \, dx \bigr)
\\
=& W(\dot{\tilde\psi}_- , \psi_+)F_+(0)-W(\dot{\tilde\psi}_- , \psi_-)\int_0^\infty \tilde\psi_+ (x) \psi_+ (x) \, dx\,, \\
W( \frac{\partial \tilde\psi_-}{\partial z} , 
\frac{\partial \psi_+}{\partial s}) =&
F_-(0)W(\tilde\psi_- , \dot{\psi}_+)-
\Bigl(\int_{- \infty}^0 \tilde\psi_- (x) \psi_- (x) \, dx\Bigr)
W(\tilde\psi_+ , \dot{\psi}_+)\,, 
\end{align*}
Taking the trace of difference of the two expressions, the first terms on 
the r.h.s. cancel because of (\ref{eq:dc1}, \ref{eq:dc2}). The 
result is that $C$ agrees with the r.h.s. of (\ref{eq:QT3}). 

The stated independence of the trace follows from its cyclicity by joining the
left and right actions (\ref{eq:ra}) in such a way as to preserve 
(\ref{eq:nrmlzt}); that of the integral is explained after Eq.~(\ref{eq:C}).
\hfill$\blacksquare$\\

\noindent 
{\bf Proof of Theorem~\ref{thm:2}.}
i) We recall that the scattering matrix $S_L=
\bigl(\begin{smallmatrix} R_L&T'_L\\T_L&R'_L\end{smallmatrix}\bigr)$ is
that of the potential truncated to the interval $[0,L]$.
The left incident solution of (\ref{eq:psi}) is given by the expressions
(\ref{eq:incident}) in the intervals $x\le 0$, resp. $x\ge L$. Its adjoint is
a solution of (\ref{eq:chi}) since $z=\mu$ is real. By the constancy of the
Wronskian, 
\begin{equation*}
	W( 1 \ep{-\I kx} + R_L^* \ep{\I kx} , \psi_\pm;x=0) = 
W(T_L^* \ep{-\I kx}, \psi_\pm;x=L) \, ,
\end{equation*}
and by $W(1 \ep{ \I kx},\psi_\pm;x) = 
\ep{\I kx} (\psi_\pm'(x) - \I k\psi_\pm(x))$
we find
\begin{equation}\label{eq:sl}
    \bigl(\psi_\pm'(0) + \I k\psi_\pm(0)\bigr) +
R_L^* \bigl(\psi_\pm'(0)-\I k\psi_\pm(0)\bigr)=
T_L^*\ep{-\I kL}\bigl(\psi_\pm'(L)+\I k\psi_\pm(L)\bigr) \, .
\end{equation}
We have that
\begin{align}
\label{eq:lim1}
\lim_{x\to +\infty}\psi_+'(x)+\I k\psi_+(x)&=0\,,\\
\label{eq:lim2}
\lim_{x\to +\infty}\bigl(\psi_-'(x)+\I k\psi_-(x)\bigr)^{-1}&=0\,.
\end{align}
Indeed, the first limit just repeats the definition (\ref{eq:psi+}) and the
second may be rephrased to the effect that 
\begin{equation*}
A(x)\DEF
\bigl(\psi_-'(x)+\I k\psi_-(x)\bigr)^*\bigl(\psi_-'(x)+\I k\psi_-(x)\bigr)
\end{equation*}
is invertible with $\lim_{x\to +\infty}\|A(x)^{-1}\|=0$. We note that 
\begin{equation*}
A(x)=\psi_-'(x)^*\psi_-'(x)+k^2\psi_-(x)^*\psi_-(x)\,, 
\end{equation*}
since the cross term
is $-\I kW(\psi_-^*,\psi_-)=0$ by (\ref{eq:nrmlzt3}). If the claim were false,
there would exist a sequence $x\to\infty$ and $a(x)\in\mathbb{C}^n$, 
($\|a(x)\|=1$) such that $\|\psi_-'(x)a(x)\|+\|\psi_-(x)a(x)\|$
remains bounded. Together with (\ref{eq:lim1}) this however contradicts 
the fact that $W(\psi_+^*, \psi_-)$ is regular. Having so established 
(\ref{eq:lim2}), we multiply the $-$ version of (\ref{eq:sl}) by 
$\ep{\I kL}(\psi_-'(L)-\I k\psi_-(L))^{-1}$ from the right, while keeping 
the $+$ version unchanged. As $L\to+\infty$
the two equations then go over to 
\begin{align}\label{eq:s}
\bigl(\psi_+'(0) + \I k\psi_+(0)\bigr) +
R^* \bigl(\psi_+'(0)-\I k\psi_+(0)\bigr)&=0\,,\\
0&=T^*\,, \nonumber
\end{align}
in the sense that the coefficients do. Since the latter system has a unique
solution $(R^*, T^*)$, it is the limit of $(R_L^*, T_L^*)$.\\

\noindent
ii) As indicated at the end of Section~\ref{section:2}, part (ii) is an
immediate consequence of Lemma~\ref{mainlm}.
\rightline{$\blacksquare$}\\

As a preliminary to the proof of Lemma~\ref{mainlm}(i) we state: 
\begin{lm}\label{lm:prlm}
Let $\psi_+\in S^+_{(z,s)}$ and $x\in \mathbb{R}$. Then 
$0$ is an eigenvalue of $\psi_+(x)$ iff $z$ is a Dirichlet eigenvalue for 
$H(s)$ on $[x,\infty)$, including multiplicities. These conditions can 
occur only for $z\in \mathbb{R}$ and for isolated $x$.
\end{lm}
\noindent
{\bf Proof.} Solutions $\varphi=\varphi(x)$ with values in $\mathbb{C}^n$ 
of the differential equation $H(s)\varphi=z\varphi$ are square-integrable at
$x=+\infty$ iff $\varphi(x)=\psi_+(x)a$ for some $a\in \mathbb{C}^n$. Hence the
equivalence of the two conditions.  They
imply $z\in \mathbb{R}$ because the operator $H(s)$ with Dirichlet 
boundary conditions on $[x,\infty)$ is self-adjoint. To show
that $x$ is isolated, we assume $x=0$ without loss and Taylor expand 
$\psi_+(x)$ at
$x=0$ up to second order. Using (\ref{eq:psi}) on the second derivative, we 
so obtain
\begin{multline*}
\psi_+(x)^*\psi_+(x)=\\
P^\perp\bigl(\psi_+(0)^*\psi_+(0)+
x(\psi'_+(0)^*\psi_+(0)+\psi_+(0)^*\psi'_+(0))+x^2\psi_+(0)^*(V(0)-z)\psi_+(0)\bigr)P^\perp+\\
x^2\psi'_+(0)^*\psi'_+(0)+o(x^2)\,,\qquad
(x\to 0)\,,
\end{multline*}
where an orthogonal projection $P^\perp=1-P$ onto 
$(\ker \psi_+(0))^\perp$ has been
inserted for free as a result of $\psi_+(0)P=0$ and of 
$\psi'_+(0)^*\psi_+(0)=\psi_+(0)^*\psi'_+(0)$, which follows from 
(\ref{eq:refl}) for $\bar z=z$. For small $x\neq 0$ the two terms are positive
semidefinite, with the first one being definite on $(\ker \psi_+(0))^\perp$. 
Since 
\begin{equation}\label{eq:dn}
\ker \psi_+(0)\cap\ker \psi'_+(0)=\{0\}
\end{equation}
by (\ref{eq:regular}), their sum is 
positive definite on all of $\mathbb{C}^n$. Hence $\psi_+(x)$ is regular.
\hfill$\blacksquare$\\

\noindent
{\bf Proof of Lemma~\ref{mainlm}.} We keep $x_0=0$ throughout the proof.\\

\noindent
i) If at $(z_*,s_*)$ a matrix $\psi_+(0)$ 
is singular, that remains true under gauge transformations (\ref{eq:ra}). By 
the previous lemma, 
$z_* \in \gamma$ is real and not below the spectrum of $H(s_*)$. 
It remains to prove the properties holding true for a dense set of 
potentials. Eigenvalue curves $f(s)$ of 
the Dirichlet Hamiltonian $H(s)$ on $[0,\infty)$ are continuously 
differentiable, even through crossings. By Sard's theorem the set 
$\{\mu'\in\mathbb{R}\mid f(s_*)=\mu', f'(s_*)=0 \text{ for some 
$s_*\in S^1$}\}$ has zero measure. Upon adding to $V(x,s)$ an arbitrarily 
small constant we may assume that $\mu$ is not in that set. In
particular, the points $s_*$ are isolated, as claimed. We further perturb 
$V$ by 
$t W(x,s)$ where $t$ is small and $W=W(x,s)$ is an arbitrary Hermitian matrix  
from the same class as $V$. To first order in $t$, the splitting of a degenerate Dirichlet
eigenvalue $\mu$ of $H(s_*)$ is $\mu+t\tilde \mu+o(t^2)$, ($t\to 0$), where
the $\tilde \mu$ are obtained by solving the finite dimensional eigenvalue 
problem 
\begin{equation}\label{eq:evlp}
P\Bigl(\int_0^\infty dx\, \psi_+(x)^*W(x,s_*)\psi_+(x)\Bigr)P a=
\tilde \mu P\Bigl(\int_0^\infty dx \,\psi_+(x)^*\psi_+(x)\Bigr)P a\, ,
\qquad (a \in \mathbb{C}^n)\,,
\end{equation}
and $P$ is again the projection onto $\ker \psi_+(0)$. Since $\psi_+(x)$ is 
regular a.e., the matrix in brackets on the l.h.s. may take arbitrary Hermitian
values, while that on the r.h.s. is positive definite on $\mathbb{C}^n$; the 
latter may then be set equal to $1$ by means of a
gauge transformation. As a result, the eigenvalues $\tilde \mu$
are generically distinct and, since $f'(s_*)\neq 0$, the points $s_*$ split
into non-degenerate ones. Moreover, points $s_*$ with
$\det\psi'_+ (x_0)=0$ correspond to Neumann eigenvalues. They are also
perturbed and split according to (\ref{eq:evlp}), except that $P$ now is the
projection onto $\ker \psi'_+(0)$. Because of (\ref{eq:dn}) the coincidence 
between Dirichlet and Neumann eigenvalues is generically lifted.\\

\noindent
ii) If $\psi_{(z,s)}(x)$ is a solution of (\ref{eq:psi}), then $\psi_{(\bar z,s)}(x)^*$ is a solution of (\ref{eq:chi}). Hence
\begin{equation*}
  L(\bar z,s)^*-L(z,s)=W(\psi_{(\bar z,s)}^*,\psi_{(z,s)};0)=0,
\end{equation*}
by (\ref{eq:nrmlzt3}), proving the reflection property. The statement about 
the eigenvalue branch follows from (i). The winding number can be read off 
from the linearization 
\begin{equation*}
\lambda(z,s)=\frac{\partial\lambda}{\partial z}\Big|_{(\mu,s_*)}\cdot(z-\mu)+
\frac{\partial\lambda}{\partial s}\Big|_{(\mu,s_*)}\cdot(s-s_*)
+O(|z-\mu|^2+|s-s_*|^2)\,,
\end{equation*}
where the derivatives are real. \\

\noindent
iii) In view of the right action (\ref{eq:ra}) a section 
$\psi^0_+:(z,s) \mapsto \psi^0_{(z,s)}(x)$ may be defined on all of
the torus by $\psi^0_{(z,s)}(0)=1$, except for the points $(\mu, s_*)$ of 
part (i). We use it outside
of the union $\cup_{s_*}U_{s_*}$ of arbitrarily small
neighborhoods of those points; inside we use a section 
$\hat{\psi}_+$ defined there. Using these local sections, the connection
is expressed as a 1-form on the corresponding patches of the torus, e.g.
$\psi^{0*}_+\mathcal{A}$ (with ${}^*$ exceptionally denoting the pull-back), 
and the trace of the curvature as a 2-form, 
$\tr D\mathcal{A}=d\tr \psi^{0*}_+\mathcal{A}$. Upon changing the patch we 
have $\hat{\psi}_+=\psi^0_+T$ with $T=T(z,s)\in\GL(n)$ and hence
$\hat{\psi}^*_+\mathcal{A}=T^{-1}(\psi^{0*}_+\mathcal{A})T+T^{-1}(dT)$. 
So, using Stokes' theorem on (\ref{eq:C}), we express the Chern number as
\begin{equation*}
C=\frac{\I}{2\pi}\sum_{s_*} \oint_{\partial U_{s_*}}
\tr\hat{\psi}^*_+\mathcal{A}-\tr\psi^{0*}_+\mathcal{A}=
\frac{\I}{2\pi} \oint_{\partial U_{s_*}}d\log\det T\,.
\end{equation*}
We may here replace $T=\hat{\psi}_{(z,s)}(0)\psi^0_{(z,s)}(0)^{-1}=
\hat{\psi}_{(z,s)}(0)$ by $L(z,s)$, because of (\ref{eq:reg}). In 
$U_{s_*}$ we have $L(z,s)=\lambda(z,s)P(z,s)+\tilde L(z,s)$, 
where $P(z,s)$ is a rank 1 projection and $\tilde L(z,s)$ is a regular linear
map from $\ker P(z,s)$ to itself. Thus $\det L$ can be in turn replaced by 
$\det(\lambda P)=\lambda$ and the claim follows. \\

\noindent
iv) Let $u \in \mathbb{C}^n$ be the normalized eigenvector of $L(\mu,s_*)$ 
with eigenvalue $\lambda(\mu,s_*)=0$. Then 
\begin{equation}\label{eq:nove}
    \frac{\partial \lambda}{\partial z}\big|_{(\mu,s_*)} = 
\bigl( u, \frac{\partial L}{\partial z}\big|_{(\mu,s_*)} u \bigr)
=\bigl( u, \, {\psi_+^{'*}} 
\frac{\partial \psi_+}{\partial z} u \bigr) \, ,
\end{equation}
since $\psi_+ u =0$ at $(z=\mu,s=s_*)$. There we may write
\begin{equation*}
    \frac{\partial \lambda}{\partial z} = 
\bigl( u,  ({\psi_+^{*}}' \frac{\partial \psi_+}{\partial z} - \psi_+^* \frac{\partial^2 \psi_+}{\partial x \partial z})u \bigr)=  - \bigl( 
u, \, W(\psi_+^*, \frac{\partial \psi_+}{\partial z}; x=0) u \bigr)\, .
\end{equation*}
On the other hand we have
\begin{equation*}
    W(\psi_+^*, \frac{\partial \psi_+}{\partial z};x) = 
\int_x^{\infty} dx' \psi_+^* (z,x')\psi_+(z,x') > 0 \, ,
\end{equation*}
which follows by differentiating (\ref{eq:wr}) w.r.t. $x$ and by 
using (\ref{eq:psi}).\\

\noindent 
v) The matrix $R$ in (\ref{eq:cinque}) is determined by (\ref{eq:s}) or, 
after multiplication with $R$,
\begin{equation*}
    R\bigl(\psi_+'(0) + \I k\psi_+(0)\bigr) + \bigl(\psi_+'(0)-\I k\psi_+(0)\bigr)=0 \, .
\end{equation*}
This shows that $\psi_+(0)$ has eigenvalue 0 iff $R$ has eigenvalue $-1$: $\psi_+(0) u =0$ implies $(R+1) \psi_+'(0) u =0$; 
conversely $(R+1)v=0$ implies $R^*v=-v$ and then $\psi_+^*(0) v = 0$. Moreover
\begin{equation}\label{eq:dieci}
    \dot{R} \bigl(\psi_+'(0)+\I k\psi_+(0)\bigr)+R\bigl(\dot{\psi}'_+(0)
+\I k \dot{\psi}_+(0)\bigr) + \dot{\psi}'_+(0) - \I k \dot{\psi}_+(0)=0 \, .
\end{equation}
We compute the rate at which the eigenvalue crosses $-1$ as
\begin{equation*}
    \dot{Z} = \frac{\bigl(\psi_+'(0)u, \, \dot{R} \psi_+'(0)u\bigr)}{\bigl(\psi_+'(0)u, \, \psi_+'(0)u\bigr)} \, ,
\end{equation*}
since the eigenprojection of the unitary $R$ is orthogonal. Multiplying (\ref{eq:dieci}) with $\psi_+'(0) u$ from the left and with 
$u$ from the right we obtain, using $R^* \psi_+'(0)u=-\psi_+'(0)u$,
\begin{equation*}
    \bigl( \psi_+'(0)u, \, \dot{R} \psi_+'(0)u\bigr)-2\I k\bigl(\psi_+'(0)u, \, \dot{\psi}_+(0)u\bigr)=0
\end{equation*}
and hence
\begin{equation*}
    \dot{Z} \, \bigl( \psi_+'(0)u, \, \psi_+'(0) u \bigr) = 2 \I k \frac{\partial \lambda}{\partial s} \, .
\end{equation*}
\rightline{$\blacksquare$}

\appendix

\section{Adiabatic evolution}

We consider the usual quantum mechanical, adiabatic setting in presence of a spectral gap: A family of operators $H(s)$
depending smoothly on $s$ and corresponding spectral projections $P_0(s)$ belonging to an interval $I(s)$ whose endpoints lie in 
the resolvent set $\rho(H(s))$. Let $U_\varepsilon (s,s_0)$ be the propagator for the non-autonomous Hamiltonian 
$H(s)$ with $s = \varepsilon t$. Then
\begin{equation*}
        U_\varepsilon (s, s_0) (P_0(s_0)+ \varepsilon P_1(s_0)) U_\varepsilon (s, s_0)^* = P_0(s) + \varepsilon P_1 (s) + O (\varepsilon^2 )\,, \qquad 
(\varepsilon \rightarrow 0)
\end{equation*}
with $P_1 (s)$ as given by Eq.~(\ref{eq:Puno}). This result is implicit in 
\cite{Th}. We give an alternate derivation which does not approximate the 
continuous spectrum by a quasi-continuum of discrete eigenvalues. \\

\noindent
{\bf Proof.} In Eq.~(\ref{eq:Pzero}) $P_1(s)$ is uniquely 
determined \cite{Nen} by the conditions 
\begin{equation}\label{eq:conds}
\begin{gathered}
\I{\dot P}_0(s)=[H, P_1 (s)]\,,\\
P_0(s)P_1 (s)+P_1 (s)P_0(s)=P_1 (s)\,,
\end{gathered}
\end{equation}
which are obtained by differentiating the expansion w.r.t. $s$, respectively 
from the fact that it represents a projection. We omit $s$ from the notation 
in the rest of the proof. 
Eq.~(\ref{eq:Puno}) satisfies the first condition because of
\begin{equation*}
[H, P_1]= -\frac{1}{2\pi}\oint_{\gamma}  [H-z, R(z) \dot{R} (z)]dz
=-\frac{1}{2\pi}\oint_{\gamma} 
(\dot{R} (z) +R(z)^2\dot{H})dz\,,
\end{equation*}
where we expanded the commutator and used $\dot{R}=-R\dot{H}R$. The second
contribution vanishes and the first yields the claim by 
$P_0= -(2\pi \I)^{-1} \oint_\gamma R(z) \, dz$.
The second condition (\ref{eq:conds}) is equivalent to $P_0P_1P_0=0$, 
$(1-P_0)P_1(1-P_0)=0$, 
which are satisfied, too: we rewrite $\dot{R}$ as before and 
use the spectral representation 
$P=\int_I dP_\lambda$ to compute
\begin{equation*}
   P_0P_1P_0=  \int_    I  \int_I (dP_\lambda) \dot{H} (d P_\mu) 
\oint_\gamma dz \frac{1}{(\lambda - z)^2(\mu - z)} = 0 \, ;
\end{equation*}
similarly, $(1-P_0)P_1(1-P_0)=0$.\hfill$\blacksquare$\\

We may add that in \cite{ASY}, Eq.~(2.6) and \cite{ASY2}, Eq.~(2.10a), as well
as in \cite{Nen}, Eq.~(2.28), the expression 
\begin{equation}\label{eq:asy}
    P_1 (s) = -\frac{1}{2\pi} \oint_{\gamma (s)} R(z,s) [\dot{P}(s),P(s)] R(z,s) \, dz
\end{equation}
is given. Its equality with (\ref{eq:Puno}) can be verified independently of
(\ref{eq:conds}).

\medskip
\noindent
{\bf Acknowledgments.} We thank M.~B\"uttiker for drawing our attention to 
the multi-channel case, and S. Jansen and R. Seiler for discussions. We are
grateful to the referee who contributed the above derivation of 
Eq.~(\ref{eq:Puno}) in replacement of a longer one.

\end{document}